\documentclass{article}  
\usepackage{padre2008}
\usepackage{graphicx}

\newcommand{\qhatave}{$\langle$\^{q}$\rangle$}
\newcommand{\pvalue}{p-value}

\frompage{000} \topage{000}                                              
\def\rightmark{The Letter Q}
\def\leftmark{J.L. Nagle}
 
\title{The Letter Q (and Quantitative Constraints on the QGP)} 
\authors{
{J.L. Nagle$^1$ %
}\\[2.812mm]
{\normalsize
\hspace*{-8pt}$^1$ University of Colorado, \\ 
Boulder, CO 80309-0390\\[0.2ex] 
}}
 
\abstract{In this proceedings, we briefly review the methodology from~\cite{ppg079}
for quantitatively constraining theoretical model parameters from experimental measurements
including statistical and systematic uncertainties.  We extend this methodology to 
additional parton energy-loss calculations for single inclusive high $p_T$ particle suppression,
and also extend the comparisons to di-jet observations.  This is only the start of a 
process to give quantitative constraints on the quark-gluon plasma, and substantial theoretical
uncertainties need to be reduced/resolved in a parallel path.
}
 
\begin{document}
 
\maketitle
\setcounter{page}{1}

\section{Introduction}\label{intro}

The field of relativistic heavy ion physics is undergoing a necessary transition from 
a field focused on the declaration of discovery - of thermalized partonic matter or the 
quark-gluon plasma, to one focused on the quantitative understanding of the unique properties
of the medium created.  This effort requires a more sophisticated and full treatment of experimental
uncertainties, in particular systematic uncertainties and their underlying correlations.  Additionally,
the effort requires the reconciliation of variant theoretical approaches, or the discarding of some
in favor of others.  This proceedings focuses primarily on the first effort, though with some
commentary on the second.  

\section{Single Hadron High $p_T$ Suppression}

\begin{figure}[htb]
\vspace*{-.0cm}
\begin{center}
\includegraphics[scale=0.6]{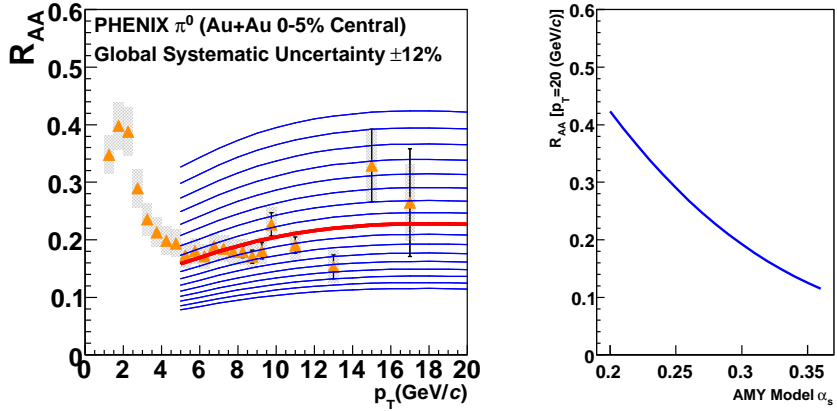}
\vspace*{-.8cm}
\end{center}
\caption[]{
(Color Online) The left panel shows the PHENIX $\pi^{0}$ nuclear modification factor as a function of $p_T$.  The curves are
different calculations of the AMY+Hydro model corresponding to different input $\alpha_s$ values.  The right
panel shows the correlation between the input $\alpha_s$ coupling and the predicted nuclear modification factor.
}
\label{fig_amy_compare}
\end{figure}

\begin{figure}[htb]
\vspace*{-.0cm}
\begin{center}
\includegraphics[scale=0.4]{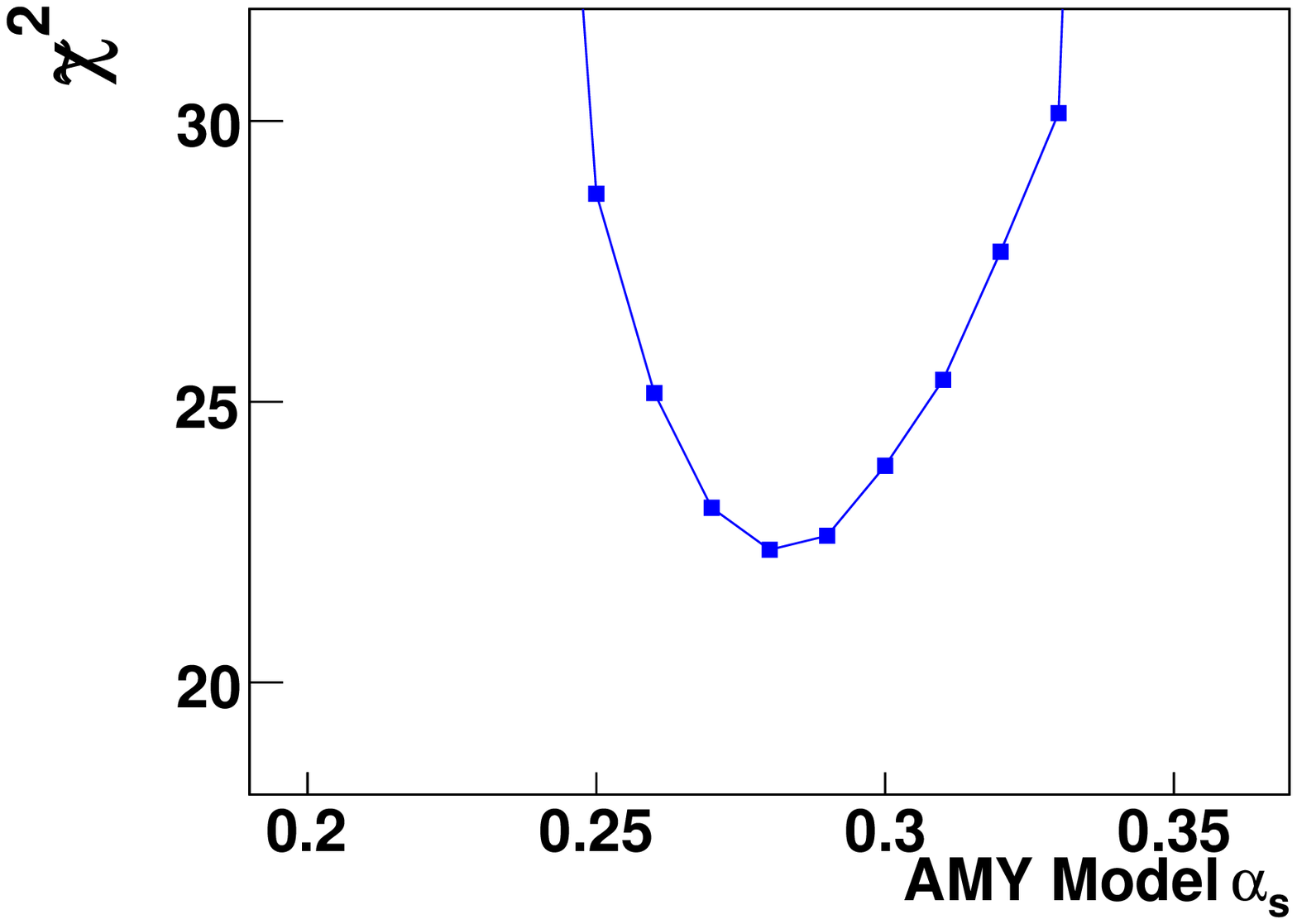}
\vspace*{-.8 cm}
\end{center}
\caption[]{(Color Online) The modified $\chi^2$ distribution as a function of the AMY+Hydro model input parameter $\alpha_s$.
}
\label{fig_amy_constraint}
\end{figure}

High $p_T$ $\pi^{0}$'s are suppressed in heavy ion reactions~\cite{ppg080}, as shown in the left panel of
Fig.~\ref{fig_amy_compare}.  The experimental
uncertainties are separated into three categories.  Type A uncertainties are point-to-point
uncorrelated and are shown as standard error lines.  For this measurement, they are dominated
by statistical uncertainties.  Type B uncertainties are point-to-point
correlated and are shown as grey bars.  They are dominated by the energy scale uncertainty
in the Electro-magnetic calorimeters and have some contribution for $p_{T} \approx 15-20$ GeV/c from
photon shower merging effects.  Type C uncertainties are globally correlated (i.e. all points move by a 
common multiplicative factor).  Note that this does {\bf not} mean that all points move up and down together
on a linear y-axis plot.  This contribution is quoted as a $\pm$ 12\% global 
systematic uncertainty and has roughly equal contributions from the uncertainty in the nuclear \
thickness function $T_{AA}$ (as determined by a Glauber model) and from the proton-proton inelastic cross
section absolute normalization.  It is notable that these last two sources of uncertainty may prove quite
difficult to reduce in the future.

These Type B and C uncertainties correspond to one RMS deviations about the measured value.  
In the analysis that follows, these uncertainties
are assumed to follow a Gaussian distribution with the corresponding RMS value.  It is notable that if
there are many contributions to the uncertainties, the Central Limit Theorem makes this assumption more solid.
However, if the best fit results are a few standard deviations away in the systematic uncertainties, one should
view the exact numerical fit result with some skepticism.  Of course, if this is the case, the best fit typically is
a very poor fit. It is also notable that if systematic uncertainties are quoted as 'full extent', then they cannot
be added in quadrature and preserve their 'full extent' nature.

Every publication of data in the field on which a full quantitative analysis is to be performed needs to
explicitly quote these RMS uncertainty contributions in their appropriately labeled category.

The methodology is detailed in~\cite{ppg079}, and here we just summarize the key result.  One calculates a modified
$\chi^2$ for a given data set with points ($y_{i}$) and a given theoretical model with predictions $\mu_{i}$ 
as a function of parameter set $p$.  One includes the possibility of systematic offsets given by the number
of standard deviations for Type B ($\epsilon_b$) and for Type C ($\epsilon_c$):
\begin{equation}
\tilde{\chi}^2={\left[\sum_{i=1}^{n}
{{(y_i+\epsilon_b \sigma_{b_i} +\epsilon_c y_i \sigma_c -\mu_i(\vec{p}))^2}  \over {{\tilde{\sigma}}_i^2}}+ {\epsilon_b^2 }+{\epsilon_c^2 }\right]} \qquad , 
\label{eq:lstsq}
\end{equation}
where ${\tilde{\sigma}}_{i}$ is the uncertainty scaled by the multiplicative shift in $y_i$ such that the
fractional error is unchanged under shifts
\begin{equation}
\tilde{\sigma}_i=\sigma_i \left( \frac{y_i+\epsilon_b \sigma_{b_i} +\epsilon_c y_i \sigma_c}{y_i}\right) \qquad . 
\label{eq:tildesigma}
\end{equation}
\noindent
Fundamentally one is determining
if the penalty for moving the data by some number of standard deviations in a systematic uncertainty is 
compensated by an overall reduction in the modified $\chi^2$ due to an improved statistical fit.  

An example comparison with a theoretical model is shown in Fig.~\ref{fig_amy_compare}.  In this case, the $\mu_{i}$ are
the AMY+Hydro theoretical calculations~\cite{amy_hydro} as a function of the input parameters $p$, specifically being the coupling $\alpha_s$.  
In~\cite{amy_hydro}, the authors utilize the AMY formalism for parton energy-loss and simulate the underlying medium with
a hydrodynamic evolution model.  The authors assert that ``once temperature evolution is fixed by the initial
conditions and evolution [by 3+1 dimensional hydrodynamics], the coupling $\alpha_s$ is the only quantity which
is not uniquely determined.''  Shown in the left panel of Fig.~\ref{fig_amy_compare} are calculations from
this model for different input values of $\alpha_s$.  The left panel shows how the nuclear modification factor
at $p_T = 20$ GeV/c varies as a function of this input coupling value.  We have then applied the full constraint method and show the modified $\chi^2$ as a function of $\alpha_s$ in Fig.~\ref{fig_amy_constraint}.  

\begin{table}[htbp]
\begin{center}
\begin{tabular}{|l|l|c|}
\hline
Model & Constrained Medium Parameters & \pvalue \\
\hline\hline
PQM   & \qhatave~= 13.2 $^{+2.1}_{-3.2}$ [1 stdev] and  $^{+6.3}_{-5.2}$ [2 stdev] ~GeV$^{2}$/fm & 9.0\% \\
\hline
GLV   & $(dN/dy)_{gluon} = 1400 ^{+270}_{-150}$ [1 stdev] and $^{+510}_{-290}$ [2 stdev] & 5.5\% \\
\hline
WHDG  & $(dN/dy)_{gluon} = 1400 ^{+200}_{-375}$ [1 stdev] and $^{+600}_{-540}$ [2 stdev] & 1.3\% \\
\hline
ZOWW  & $\epsilon_{0} = 1.9 ^{+0.2}_{-0.5}$ [1 stdev] and  $^{+0.7}_{-0.6}$ [2 stdev] ~GeV/fm & 7.8\% \\
\hline
AMY+Hydro & $\alpha_{s} = 0.280 ^{+0.016}_{-0.012}$ [1 stdev] and $^{+0.034}_{-0.024}$ [2 stdev] & 5.0\% \\
\hline
Linear & b (intercept) = $0.168 ^{+0.033}_{-0.032}$ [1 stdev] and $^{+0.065}_{-0.066}$ [2 stdev] & 11.6\% \\
       & m (slope) = $0.0017 ^{+0.0035 }_{-0.0039}$ [1 stdev] and $^{+0.0070}_{-0.0076}$ [2 stdev] ($c$/GeV) & \\
\hline
\end{tabular}
\end{center}
\caption{Quantitative constraints on the medium parameters from various energy-loss models.}
\label{tab:result}
\end{table}

The resulting constraint for the AMY + Hydro model and for a variety of parton energy-loss calculations and a linear
functional fit are given in Table~\ref{tab:result}.  It is {\bf critical to note} that each constraint is
assuming a perfect model calculation with only one unknown parameter, i.e. the uncertainty is from experimental
sources only.

\begin{figure}[htb]
\vspace*{-.0cm}
\begin{center}
\includegraphics[scale=0.5]{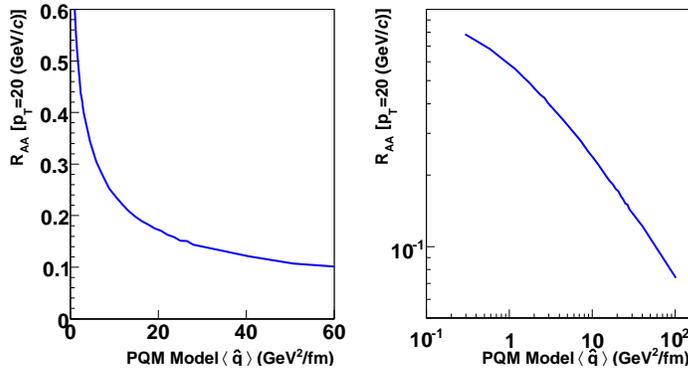}
\vspace*{-0.8cm}
\end{center}
\caption[]{
(Color Online) For the PQM energy-loss model, the correlation between the \qhatave and the nuclear modification factor is shown.
The left panel is a linear-linear display, and the right panel is a log-log display.
}
\label{fig_pqm}
\end{figure}

There are a number of general observations that can be made at this point.  In all of these calculations, there
is a clear minimum in the modified $\chi^2$ and a reasonably well defined one- and two-standard deviation limit.  
In~\cite{fragility}, the authors refer to the ``fragility of high $p_T$ hadron spectra as a hard probe.''  The usage
of this term ``fragility'' is ambiguous in the literature.  Some simply refer to it with respect to the
fact that as one increases the color-charge density or medium transport value \qhatave, the nuclear
modification factor $R_{AA}$ appears to saturate.  Thus, one can speculate that for some large value (for example
\qhatave $> 5$ GeV$^2$/fm), the uncertainty in determining $R_{AA}$ gets very large.  However, viewing such
trends on a linear y-axis scale can be quite deceptive.  Shown in Fig.~\ref{fig_pqm} are predictions from the Parton
Quenching Model (PQM)~\cite{PQM} for different \qhatave values.  In the right panel, shown on a log x-axis and log y-axis scale, is
the modification factor versus \qhatave.  It is striking that it appears linear on this
plot for \qhatave $> 5$.  Thus, for a given fractional uncertainty in measuring $R_{AA}$, one always gets the same
fractional uncertainty on \qhatave.  Very similar results are obtained with other model calculations. 

If this is not the meaning of ``fragility'' (in a purely statistical constraint sense), others refer to it 
in reference to the concept of surface emission bias.  Imagine a beam of partons aimed at the corona of a
dense medium.  
No measurement of the emitted particles provides information about the core
of the medium (since the partons are not aimed there).  However, if one had a model of the density distribution
of the medium, then measuring (for example) the color-charge density of the corona, one can (via this distribution)
learn about the density in the core.  However, this knowledge is ``fragile'' in the sense that it depends more
and more sensitively on the knowledge of the medium density distribution.  It is interesting that in the paper
discussing this ``fragility'' issue~\cite{fragility}, the authors employ a very unrealistic uniform cylindrical 
geometry, such that the density in the corona is identical to the density in the core.  It is also notable that
this geometry gets the distribution of hard scattering positions incorrect as well.

\section{Di-Jet Observables}

\begin{figure}[htb]
\begin{center}
\vspace*{-.0cm}
\includegraphics[scale=0.5]{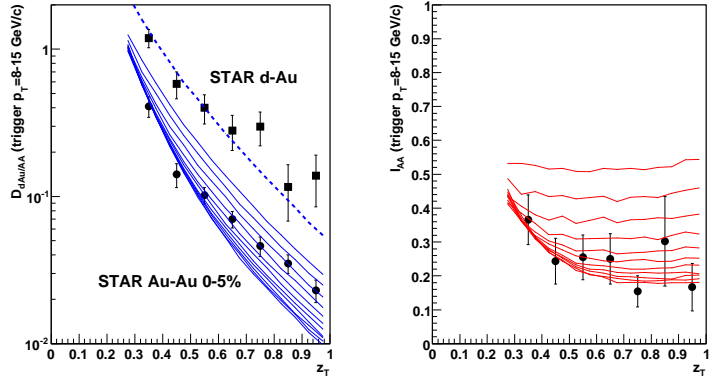}
\vspace*{-0.8cm}
\end{center}
\caption[]{
(Color Online) The left panel shows the STAR dijet for per trigger away-side integrated yield ($D_{dAu/AuAu}$) as a function of $z_T$ from
d-Au and Au-Au.  The right panel is the ratio ($I_{AA}$).  An additional 5\% and 7\% Type C uncertainty is including in the
$D_{dAu/AuAu}$ and $I_{AA}$ results, respectively~\cite{jacobs}.  Also shown are the ZOWW theory calculations for different
input energy-loss parameter ($\epsilon_{0}$) values.
}
\label{fig_star_dijet}
\end{figure}

\begin{figure}[htb]
\begin{center}
\vspace*{-.0cm}
\includegraphics[scale=0.3]{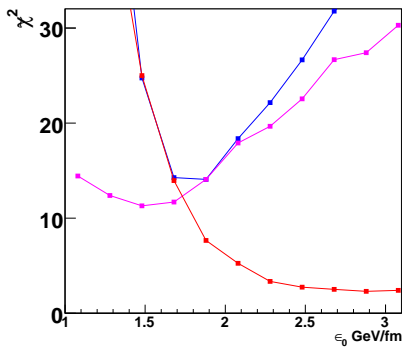}
\vspace*{-0.8cm}
\end{center}
\caption[]{(Color Online) The modified $\chi^2$ for the ZOWW calculations for the $I_{AuAu}$ results (red line), the $D_{AuAu}$ only (blue line),
and the $D_{AuAu}$ including the theoretical scale uncertainty (magenta line).
}
\label{fig_dijet_constraint}
\end{figure}

Regardless of the definition of ``fragility'', it is always wise to search for additional discriminating
experimentally accessible observables.  In~\cite{zoww}, they speculate that di-jet observables will be more
sensitive to medium parameters.  In the context of their calculation (ZOWW), they show that $I_{AA}$ (the modification
of per trigger away-side yields) has a steeper dependence on their energy-loss parameter $\epsilon_0$ (GeV/fm),
than the single inclusive modification factor $R_{AA}$.  Thus, if one had identical experimental uncertainties
on the two quantities, then $I_{AA}$ would provide the stricter constraint.  

Shown in Fig.~\ref{fig_star_dijet} are the experimental data from the STAR experiment~\cite{star_dijet}, and the
calculations from the ZOWW model~\cite{zoww}.  If we apply a 7\% Type C uncertainty to the STAR $I_{AA}$ 
results~\cite{jacobs}, the resulting constraint is $\epsilon_{0} = 2.9 ^{+???}_{-0.6}$ [one std. dev.] and $^{+???}_{-0.9}$
[two std. dev.].  The '???' refers to the fact that within the parameter ranges available, there is no upper constraint.
This is notably worse than the constraint from $R_{AA}$ given in Table~\ref{tab:result}.  In fact,
the two results do not overlap each other within the one standard deviation uncertainties.  The worse constraint
from $I_{AA}$ is due to the poor statistics in the $d-Au$ reference (used instead of proton-proton) for the
modification factor.  This will be remedied in the near future with the much larger $d-Au$ data set taken in 2007-2008.

It is notable that in~\cite{zoww}, they show a rather tight $\epsilon_{0}$ constraint in a figure labeled $I_{AA}$.
However, the actual constraint was derived from just comparing the Au-Au result alone ($D_{AuAu}$).  Thus, the constraint
is only derived from the Au-Au data in the left panel of Fig.~\ref{fig_star_dijet}.  The resulting modified $\chi^2$'s 
from $I_{AA}$ and $D_{AuAu}$ alone are shown in Fig.~\ref{fig_dijet_constraint}.  It is notable that the fit to the $D_{AuAu}$ alone
is a rather poor fit, which can be visually seen in Fig.~\ref{fig_star_dijet} where the shape of the data points for $z_T > 0.4$ 
do not match any of the theory curves.  If one only fits $D_{AuAu}$, there is a definite additional theory uncertainty from
the pQCD scale uncertainty in the NLO calculation (shown in~\cite{zoww}).  If this is included, the constraint is much looser, again
shown in Fig.~\ref{fig_dijet_constraint}.  Future high statistics data sets will allow for a more detailed study.  Also, the
use of the variable $z_T$ may hide many details, and an optimal high statistics presentation should first include fine binning
in trigger $p_T$ for $I_{AuAu}$.

\section{Summary}

We summarize with a set of observations.  On the experimental side, we have a well understood
method for inclusion of statistical and systematic uncertainties.  Experiments need to carefully
quantify these Type A, B, and C uncertainties for all relevant measurements.  It will be 
very interesting to see if larger p-p and d-Au data sets reconcile the $R_{AA}$ and $I_{AA}$ constraints.

On the theoretical side, one needs to resolve fundamental disconnects about whether perturbative
calculations with modest $\alpha_s$ fully describe parton energy loss.  Currently the PQM results~\cite{PQM}
indicate that the perturbative calculations fail.  In either case, this must be reconciled at the appropriate
scales with the picture of the bulk medium being strongly coupled (near-perfect fluid) and not being
describable in a perturbative framework.  Also, all calculations {\bf must}
include realistic geometries, fluctuations, and running of the coupling (if possible) 
so that the discussions and comparisons can focus on the more fundamental physics questions.

\section*{Acknowledgments}
The author acknowledges fruitful collaboration with M.J. Tannenbaum, and useful
theory and experiment input and discussions with W. Horowitz, P. Jacobs, C. Loizides, G-Y Qin,
I. Vitev, and X.N. Wang. 
The author also acknowledges
funding from the Division of Nuclear Physics of the U.S. Department of Energy under Grant No.
DE-FG02-00ER41152.

\vfill\eject
\end{document}